\documentclass{ctr_summer}

\usepackage{ctrfont}
\usepackage{natbib}
\usepackage{undertilde}
\usepackage{color}

\usepackage{graphicx}

\usepackage{amsmath,rotating}
\usepackage{dashrule}
\usepackage{undertilde}

\usepackage{url}

\usepackage{color}

\newcommand{\mol}[1]{\ensuremath{\underline{#1}}}
\newcommand{\vect}[1]{\ensuremath{\mathbf{#1}}}
\providecommand\bnabla{\boldsymbol{\nabla}}
\providecommand\bcdot{\boldsymbol{\cdot}}

\usepackage{array}
\newcolumntype{P}[1]{>{\centering\arraybackslash}p{#1}}

\urldef{\ECNmovies}\url{http://www.sandia.gov/ecn/cvdata/assets/movies/bkldaAL1movie.php} 
\urldef{\ECNliq}\url{http://www.sandia.gov/ecn/cvdata/assets/datafiles/liq/bkldaAL1liquid.php} 
\urldef{\ECNvap}\url{http://www.sandia.gov/ecn/cvdata/assets/datafiles/pen/bkldaAL1-pen.txt} 
\urldef{\ECNmixture}\url{http://www.sandia.gov/ecn/cvdata/assets/Rayleigh/bkldaAL4mixing.php} 
\urldef{\ECNmassflowrate}\url{http://www.cmt.upv.es/ECN03.aspx} 
\urldef{\ECN}\url{http://www.sandia.gov/ecn/} 

\newcommand{\gnult}[1]{%
	\ifcase#1%
	\or 
		\mbox{( \solid\ )}%
	\or 
		\mbox{( \dashed\ )}%
	\or 
		\mbox{( \sdashed\ )}%
	\or 
		\mbox{( \dotted\ )}%
	\or 
		\mbox{( \dashdot\ )}%
	\or 
	\or 
		\mbox{( \dashdash\ )}%
	\or 
		\mbox{( \dashdotdot\ )}%
	\else%
	\fi%
}%

\newcommand{\gnultc}[2]{%
	\ifcase#1%
	\or 
		\mbox{( \textcolor{#2}{\solid} )}%
	\or 
		\mbox{( \textcolor{#2}{\dashed} )}%
	\or 
		\mbox{( \textcolor{#2}{\sdashed} )}%
	\or 
		\mbox{( \textcolor{#2}{\dotted} )}%
	\or 
		\mbox{( \textcolor{#2}{\dashdot} )}%
	\or 
	\or 
		\mbox{( \textcolor{#2}{\dashdash} )}%
	\or 
		\mbox{( \textcolor{#2}{\dashdotdot} )}%
	\else%
	\fi%
}%

\newcommand{\gnupt}[1]{%
	\ifcase#1%
	\or 
		\mbox{( $+$ )}%
	\or 
		\mbox{( $\times$ )}%
	\or 
	\or 
		\mbox{( $\boxdot$ )}%
	\or 
		\mbox{( {\scriptsize $\blacksquare$} )}%
	\or 
		\mbox{( $\odot$ )}%
	\or 
		\mbox{( {\scriptsize $\CIRCLE$} )}%
	\or 
		\mbox{( {\footnotesize $\vartriangle$} )}%
	\or 
		\mbox{( {\footnotesize $\blacktriangle$} )}%
	\or 
		\mbox{( {\footnotesize $\triangledown$} )}%
	\or 
		\mbox{( {\footnotesize $\blacktriangledown$} )}%
	\or 
		\mbox{( {\scriptsize $\lozenge$} )}%
	\or 
		\mbox{( {\scriptsize $\blacklozenge$} )}%
	\or 
		\mbox{( $\pentagon$ )}%
	\else%
	\fi%
}%

\usepackage{dashrule}

\newcount\ndots
\def\drawline#1#2{\raise 2.5pt\vbox{\hrule width #1pt height #2pt}}
\def\spacce#1{\hskip #1pt}

\def\solid{\drawline{24}{.6}\nobreak}

\def\dashed{\hbox{\drawline{4.5}{.6}\spacce{2}\drawline{4.5}{.6}\spacce{2}\drawline{4.5}{.6}\spacce{2}\drawline{4.5}{.6}}\nobreak}

\def\sdashed{\hbox{\drawline{2.3}{.6}\spacce{2}\drawline{2.3}{.6}\spacce{2}\drawline{2.3}{.6}\spacce{2}\drawline{2.3}{.6}\spacce{2}\drawline{2.3}{.6}\spacce{2}\drawline{2.3}{.6}}\nobreak}

\def\dotted{\hbox{\drawline{1}{.6}\spacce{1.1}\drawline{1}{.6}\spacce{1.1}\drawline{1}{.6}\spacce{1.1}\drawline{1}{.6}\spacce{1.1}\drawline{1}{.6}\spacce{1.1}\drawline{1}{.6}\spacce{1.1}\drawline{1}{.6}\spacce{1.1}\drawline{1}{.6}\spacce{1.1}\drawline{1}{.6}\spacce{1.1}\drawline{1}{.6}\spacce{1.1}\drawline{1}{.6}\spacce{1.1}\drawline{1}{.6}}\nobreak}

\def\dashdot{\hbox {\drawline{4.75}{.6}\spacce{2}\drawline{1}{.6}\spacce{2}\drawline{4.5}{.6}\spacce{2}\drawline{1}{.6}\spacce{2}\drawline{4.75}{.6}}\nobreak}

\def\dashdotdot{\hbox {\drawline{5}{.6}\spacce{2}\drawline{1}{.6}\spacce{2}\drawline{1}{.6}\spacce{2}\drawline{5}{.6}\spacce{2}\drawline{1}{.6}\spacce{2}\drawline{1}{.6}}\nobreak}

\def\dashdash{\hbox {\drawline{2}{.6}\spacce{1.4}\drawline{2}{.6}\spacce{3.9}\drawline{2}{.6}\spacce{1.4}\drawline{2}{.6}\spacce{3.9}\drawline{2}{.6}\spacce{1.4}\drawline{2}{.6}}\nobreak}

\title{Multi-component vapor-liquid equilibrium model for LES and application to ECN Spray A}

\shorttitle{Multi-component vapor-liquid equilibrium model for LES}

\author{J. Matheis\footnote[1]{Institute of Aerodynamics and Fluid Mechanics, Technische Universit\"at M\"unchen, Germany} \and S. Hickel\footnote[2]{Faculty of Aerospace Engineering, Technische Universiteit Delft, The Netherlands}}

\shortauthor{Matheis \& Hickel}

\begin{document}

\setcounter{page}{1}

\maketitle
We present and evaluate a detailed multi-species two-phase thermodynamic equilibrium model for large-eddy simulations (LES) of liquid-fuel injection and mixing at high pressure. The model can represent the coexistence of supercritical states and multi-component subcritical two-phase states. LES results for the transcritical Spray A of the Engine Combustion Network (ECN) are found to agree very well to available experimental data. We also address well-known numerical challenges of trans- and supercritical fluid mixing and compare a fully conservative formulation to a quasi conservative formulation of the governing equations. Our results prove physical and numerical consistency of both methods on fine grids and demonstrate the effects of energy conservation errors associated with the quasi conservative formulation on typical LES grids.  \\

\hrule


\section{Introduction}

We discuss large-eddy simulation (LES) results for the high-pressure liquid-fuel injection Spray~A benchmark case of the Engine Combustion Network (ECN, \ECN), with particular emphasis on both the physical and numerical modeling of the turbulent mixing of supercritical and transcritical fluids. The setup consists of a cold n-dodecane jet ($C_{12}H_{26} $ at $363~\mathrm{K}$) that is injected with about $600~\mathrm{m/s}$ into a warm nitrogen ($N_2$) atmosphere at $T = 900~\mathrm{K}$ and a pressure of $p = 6~\mathrm{MPa}$. 
This high pressure exceeds the critical pressure $p_c$ of both components and results in a liquid-like ($p>p_c$, $T<T_c$) and a gas-like ($T>T_c$, $p>p_c$) state of the two pure species. However, the critical pressure of certain mixtures of the two species is much higher than the critical pressure of the pure species and also higher than the Spray A operating pressure, such that the mixture locally becomes subcritical and interfaces between liquid and gas phases may appear during the mixing process. We refer to such conditions as transcritical operating conditions. 
%
%
%
Previous numerical simulations of the transcritical Spray~A have either modeled the spray with Lagrangian particle tracking (LPT) methods \citep[see, e.g.,][]{Wehrfritz:2013jr,Xue:2013ch}, i.e., as a classical two-phase spray with sharp gas-liquid interfaces (first- and secondary breakup, evaporative droplets), or with a single-phase dense-gas approach \citep[see, e.g.,][]{Lacaze:2015jt}, arguing that the high pressure and temperature lead to a miscible mixing with negligible surface tension. 
Both approaches can be justified but have obvious limitations when applied to transcritical operating conditions that correspond to a transition regime between classical spray dynamics and miscible mixing. 
Standard LPT methods are very efficient computationally, but neglect real-gas effects and dissolved ambient gases in the liquid fuel phase, which become substantial at high pressures \citep{Balaji:2011bw,Qiu:2015hc}. 
Furthermore, results can strongly depend on the values of calibration parameters. 
%
%
The single-phase dense-gas approach, on the other hand, does not include the effect of phase separation and may thus lead to unphysical or ill defined states if part of the flow is subcritical. 
To improve on these limitations, and inspired by the work of \cite{Qiu:2015hc}, we have developed a detailed multi-species two-phase thermodynamic model for the Eulerian LES of turbulent mixing at supercritical and transcritical pressures, which can represent the coexistence of multi-component subcritical two-phase states in a computational cell.

\section{Physical and numerical models}

\subsection{Governing equations}\label{Sec:GovEqn}

We solve the three-dimensional compressible multi-component Navier-Stokes equations either in a fully conservative (FC) formulation, 
\begin{gather}
	\partial_t  \rho +  \bnabla \bcdot (\rho \mathbf{u}) = 0 \label{eq:continuity} \\
	\partial_t \rho Y_i + \bnabla \bcdot ( \rho Y_i \mathbf{u}) = \bnabla \bcdot \mathbf{J}_i \label{eq:species} \\
	\partial_t \mathbf{\rho u} +  \bnabla \bcdot ( \rho \mathbf{u} \mathbf{u}  + \mathbf{I} p) = \bnabla \bcdot \boldsymbol{\tau} \label{eq:momentum} \\
	\partial_t E + \bnabla \bcdot \left[  (E + p)\mathbf{u} \right]  = \bnabla \bcdot \left( \mathbf{u} \bcdot  \boldsymbol{\tau}  - \mathbf{q} \right) \label{eq:energy} \ ,
\end{gather}
or in a quasi conservative (QC) formulation for which the total energy conservation, Eq.~\eqref{eq:energy}, is replaced by the pressure evolution equation (PEVO), c.f. \cite{Terashima:2012dd}
\begin{gather}
	\partial_t p + \bnabla \bcdot (p \mathbf{u}) = ( p - \rho c^2) \bnabla \bcdot \mathbf{u} + \frac{\alpha_p}{c_v \beta_T \rho }  \left[ \vphantom{\frac{\alpha_p}{c_v \beta_T \rho }}   \bnabla \bcdot (\boldsymbol{\tau} \bcdot \mathbf{u} -    \mathbf{q} ) - \mathbf{u} \bcdot ( \bnabla \bcdot  \boldsymbol{\tau} ) \right] \dots \nonumber\\
	\hspace{8cm} + \sum_{i=1}^N \left. \frac{1}{\rho} \frac{\partial p}{ \partial Y_i} \right |_{\rho,e,Y_j[i]} \bnabla \bcdot \mathbf{J}_i \label{eq:pevo} \ .
\end{gather}
The state vector consists of mass density $\rho$, partial densities $\rho Y_i$ of species $i = \lbrace 1 \ldots N_c \rbrace$, linear momentum $\rho \mathbf{u}$, and total energy density $ E = \rho e + \frac 1 2 \rho \mathbf{u} \bcdot \mathbf{u} $ (FC) or the pressure $p$ (QC). \mbox{$\mathbf{u} = \left[ u_1, u_2, u_3 \right]^T$} is the velocity vector in a Cartesian frame of reference, $c$ denotes the speed of sound, $c_v$ is the heat capacity at constant volume, and $\alpha_p$ and $\beta_T$ are the thermal expansion and isothermal compressibility coefficient. 
To allow for a meaningful comparison between FC and QC simulations we also included the effect of the diffusion induced pressure variation, the last term on the right hand side of Eq.~\eqref{eq:pevo}, which was neglected by \cite{Terashima:2012dd}.

According to the Stokes hypothesis for a Newtonian fluid, the viscous stress tensor is
$
	{\boldsymbol{\tau}}=  \mu \left( \bnabla \mathbf{u} +  (\bnabla \mathbf{u})^T - 2/3 \, \mathbf{I}\ \bnabla \bcdot \mathbf{u}  \right)\, ,
$
with $\mu$ being the dynamic viscosity and $\mathbf{I}$ the unit tensor. The diffusional fluxes are calculated via Fick's law 
$
	\mathbf{J}_i =  \rho D_i \bnabla Y_i - Y_i \sum_{j=1}^N  \rho D_j \bnabla Y_j  \ ,
$
where  
	$ D_i = \left( 1 - z_i  \right) / \sum_{j \ne i}^N {z_j}{D_{ij}^{-1}}  $
is an effective binary diffusion coefficient for the diffusion of species $i$ into the rest of the mixture and $z_i$ denotes the mole fraction of species $i$.  
The physical binary mass diffusion coefficients $D_{ij}$ are modeled according to Chapman and Enskog theory \citep[see e.g.][]{Prausnitz:1998vu}.
The vector 
$	\mathbf{q} = -\kappa \bnabla T\, - \sum_{i=1}^N h_i \mathbf{J}_i $
consists of heat conduction the enthalpy flux by species diffusion, where $\kappa$ is the thermal conductivity, $T$ is the temperature, and $h_i$ is the partial enthalpy of species $i$. Helpful details on the calculation of partial properties can be found \cite{Masquelet:2013tw}. Viscosity and thermal conductivity are modeled with correlations given by \citet{Chung:1988vk}. 
The FC and QC equations are closed by a thermodynamic model that relates pressure, temperature and density.



%
\subsection{Multi-component single-phase equation of state}\label{Sec:EOS}

%
Our single- and two-phase models are based on cubic equations of states (EOS) 
\begin{gather} 
	p(\mol{v},T,\vect{z}) =  \frac{\mathcal{R}T}{\mol{v} - b} - \frac{a \alpha }{ \mol{v}^2 + u\, b\, \mol{v} + w b^2 } , \label{eq:CEOS}
\end{gather}
where the pressure $p$ is a function of the molar volume $\mol{v}$, temperature $T$ and the molar composition $\vect{z} = \{z_1 \dots z_{N_c}\}$. Here and in the following, all intensive thermodynamic properties are expressed as molar quantities, denoted by $\mol{\star}$.  $\mathcal{R}$ is the universal gas constant. 

In all subsequent simulations we use the Peng-Robinson (PR) EOS \citep{Peng:1976tq} for which $u=2$ and $w=-1$. 
The function \mbox{$\alpha = [1 + c_0(1-\sqrt{T_r})]^2 $} accounts for the polarity of a fluid and is a correlation of temperature $T$, critical temperature $T_c$ and acentric factor $\omega$ via \mbox{$c_0 = 0.37464 + 1.54226\omega - 0.2699 \omega^2$}. The parameter \mbox{$a = 0.45724 \left({\mathcal{R}^2T_c^2}/{p_c} \right)$} represents attractive forces between molecules and the effective molecular volume is represented by $b = 0.0778 \left({\mathcal{R}T_c}/{p_c} \right) $. 

We use conventional mixing rules to extend the PR EOS to a mixture composed of $N_c$ components. The parameters required in the EOS are calculated from
\begin{gather}
	a\alpha = \sum_i^{N_c} \sum_j^{N_c} z_i z_j a_{ij} \alpha_{ij} \quad \text{and} \quad
	b = \sum_i^{N_c} z_i b_{i}, \label{eq:MixingRule}
\end{gather}
with $z_i$ being the mole fraction of component $i$ (overall or in the liquid/vapor phase). 
The coefficients $a_{ij}$ and $\alpha_{ij}$ are calculated with combination rules given by \citet{Harstad:1997ww}. We calculate off-diagonal elements using the same expression as for the diagonals together with pseudo-critical parameters
\begin{gather}
	T_{c,ij} = \sqrt{T_{c,i} T_{c,j}}(1-\delta^\prime_{ij}), \quad p_{c,ij} = Z_{c,ij}(\mathcal{R}T_{c,ij}/v_{c,ij}), \\[0.1cm]
	v_{c,ij} = \frac{1}{8}\left[v_{c,i}^{1/3}+v_{c,j}^{1/3}\right]^3,  \quad  \omega_{ij} = 0.5 \left(\omega_i + \omega_j\right), \quad Z_{c,ij} = 0.5 \left(Z_i + Z_j\right).
\end{gather} 
The binary interaction parameter $\delta^\prime_{ij}$ is set to zero for all simulations in this report.



Additionally to the thermal EOS, expressions for caloric properties (e.g. internal energy $\mol{e}$, specific heats $\mol{c}_p$ and $\mol{c}_v$, etc.) that account for their pressure dependance are needed. The departure function formalism provides such expressions and only requires relationships provided by the EOS, see, e.g., \citet{Poling:2000tp}. The ideal reference state is evaluated using the 9 coefficient NASA polynomials~\citep{Burcat:2005vf}. 

The single-phase frozen temperature $(T_F)$ is computed iteratively by minimizing the objective function
%
$  F^{FC} = ({\mol{e}^\star - \mol{e}_{F}(T_{F},\mol{\rho}^\star,\vect{z}^\star)})/{\mol{e}^\star} $ or $ F^{QC} = ({\mol{\rho}^\star -\mol{\rho}_{F}(T_{F},p^\star,\vect{z}^\star)})/{\mol{\rho}^\star}$,
with $\mol{e}^\star=\mol{e}_{LES}$ (FC), $p^\star=p_{LES}$ (QC), $\mol{\rho}^\star=\mol{\rho}_{LES}$ and $\vect{z}^\star=\vect{z}_{LES}$ being the molar internal energy, pressure, molar density and overall molar composition that come from the flow solver (after conversion to molar quantities).

Once the temperature is available, all other thermodynamic properties (e.g. pressure for FC formulation) and derivatives (e.g. specific heats, speed of sound, partial properties) can be calculated in a straightforward manner. It is important to note that the pressure and temperature resulting from this {single-phase model} may correspond to unstable thermodynamic states. 

\subsection{Multi-component two-phase equilibrium model}\label{Sec:EOS-M}

A mixture is considered stable at the current temperature and pressure if and only if the total Gibbs energy is at its global minimum \citep{Michelsen:2007vt}. Whether a split into two phases yields a decrease in the Gibbs energy
is determined with the Tangent Plane Distance (TPD) function \citep{Michelsen:1982vn}. 
For the present work we followed the recommendation of \citet{Qiu:2014wi} and implemented the BFGS-quasi-Newton algorithm, see \citet{Hoteit:2006ek} and references therein. If the result of the TPD test tells us that the single-phase mixture is stable, then we apply the cubic EOS in a straightforward manner. If it turns out that the mixture is unstable, which means that the fluid would prefer to exist as two phases separated by an interface, then we solve the so-called isochoric-isoenergetic flash problem. Note that this implies the assumption that the phase-transition timescale is small compared to the flow timescale.
 
Temperature and pressure are iterated until the sum (weighted by the phase fraction) of the liquid-phase and vapor-phase densities and internal energies within a computational cell corresponds to the overall internal energy and density that come from the flow solver. The corresponding objective function for the two-phase equilibrium model is
\begin{gather}
\vect{F} = 
\left \{
\frac{\mol{v}^\star - \mol{v}_{EQ}(T,p,\vect{z}^\star)}{\mol{v}^\star}, 
\frac{\mol{e}^\star - \mol{e}_{EQ}(T,p,\vect{z}^\star)}{\mol{e}^\star}
\right \}
\end{gather}
with $\mol{e}^\star=\mol{e}_{LES}$, $\mol{v}^\star=\mol{v}_{LES}$ and $\vect{z}^\star=\vect{z}_{LES}$ being the specific molar internal energy and volume and overall composition in the corresponding cell, respectively.

In the innermost iteration loop we solve an isothermal isobaric flash problem, i.e., we calculate the vapor-liquid phase equilibrium (VLE) at given temperature, pressure and overall composition. The necessary condition of thermodynamic equilibrium is that the fugacity $f_i$ of each component $i$ is the same in the liquid (subscript $l$) and vapor (subscript $v$) phase, i.e., 
$
	f_{i,v}(T,p,\vect{y}) = f_{i,l}(T,p,\vect{x})
$.
We denote liquid and vapor phase mole fractions by $\vect{x} = \{x_1 \dots x_{N_c}\}$ and $\vect{y} = \{y_1 \dots y_{N_c}\}$, respectively.
The material balance for each component,
$
	\psi_v y_i + ( 1 - \psi_v ) x_i = z_i
$,
with $\psi_v$ being the overall molar vapor fraction, and the requirement that mole fractions in the liquid and vapor phase must sum to unity, or equivalently 
$
	\sum_{i=1}^{N_c} y_i -x_i = 0,
$
yield $(2N_c+1)$ equations, which are solved for the unknown compositions $\vect{x}$ and $\vect{y}$ of liquid and vapor, and the molar vapor fraction $\psi_v$. Equilibrium volume $\mol{v}_{EQ}$ and energy $\mol{e}_{EQ}$ 
are then obtained as   
\begin{gather}
	\mol{v}_{EQ}(T,p,\vect{z}^\star) = \psi_v \mol{v}_v + (1 - \psi_v) \mol{v}_l \quad \text{and} \quad \mol{e}_{EQ}(T,p,\vect{z}^\star) = \psi_v \mol{e}_v + (1 - \psi_v) \mol{e}_l.
\end{gather}
Specific molar volumes ($\mol{v}_v(T,p,\vect{y})$,$\mol{v}_l(T,p,\vect{x})$) and energies ($\mol{e}_v(T,p,\vect{y})$,$\mol{e}_l(T,p,\vect{x})$) of the two phases are calculated with the EOS (Eq.~\ref{eq:CEOS}) and the departure function formalism, respectively.
For a comprehensive review and practical implementation guidelines the interested reader is referred to the textbook of \cite{Michelsen:2007vt}.

 \subsection{Discretization method and turbulence model}
 
The governing equations of the FC formulation, Eq.~\eqref{eq:continuity}-\eqref{eq:energy}, are discretized by a conservative finite-volume scheme. Effects of unresolved subgrid scales (SGS) are modeled by the adaptive local deconvolution method (ALDM) of  \citet{Hickel:2014tl}. In order to avoid spurious oscillations at sharp density gradients, we use the van Albada limiter 
for the reconstruction of mass and internal energy. The viscous flux is discretized with a $2^{nd}$ order central difference scheme, and a $3^{rd}$ order explicit Runge-Kutta scheme 
is used for time integration. The left hand side of the pressure evolution equation for the QC method is discretized consistently with the internal energy transport, such that both discretizations are identical up to machine precision for a single-species perfect gas.

\section{Interlude: consistency and convergence of FC and QC formulation}\label{Sec:Interlude}

\begin{figure*}
\centering
	\includegraphics[scale=0.9]{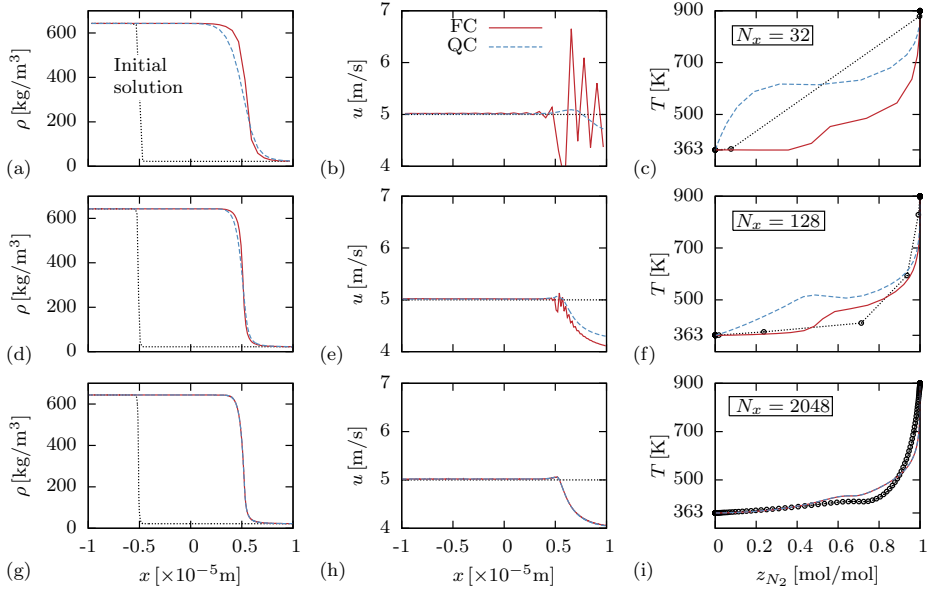}
	\caption{FC-F and QC-F results for 1-D advection-diffusion test case for different grid resolutions. Left column: density profiles in physical space; center column: velocity profiles in physical space; right column: temperature profiles in mixture space; dotted lines are the initial profiles.}
\label{fig:1D}
\end{figure*}
Equations~\eqref{eq:continuity}-\eqref{eq:energy} and \eqref{eq:continuity}-\eqref{eq:species},\eqref{eq:pevo} are expected to converge to the same solution with increasing grid resolution. To prove this important hypothesis, we show results for a 1-D advection-diffusion test case of a contact discontinuity in Figure~\ref{fig:1D}. The number of uniform cells in the region of interest ($-l_{ref}/2 < x < l_{ref}/2$) with $l_{ref}=2 \times10^{-5}$~m is varied between $32$ to $2048$ and two blocks with stretched cells are attached on both sides such that reflections from the boundary conditions cannot affect the results.

The chosen thermodynamic conditions are similar to Spray A ($p=6$~MPa, $T_{N2}=900$~K, $T_{C{12}H{26}}=363$~K) and the advection velocity is $u = 5$~m/s. Species mass fractions are initialized with an error function profile in physical space 
\begin{gather}
Y_{C12H26} = 0.5 - 0.5\;\mathrm{erf}\{( x_i + 0.25 l_{ref} ) / ( 0.01 l_{ref} ) \} 
\end{gather}
with $x_i$ being the cell-center coordinates. Both FC and QC equations are closed by the single-phase model (FC-F and QC-F). 
The temperature across the initial interface is computed from a linear enthalpy profile in mixture space, commonly known as the adiabatic mixing temperature. 

First and second columns in Figure~\ref{fig:1D} depict the density and velocity at \mbox{$t = 2\times10^{-6}$~s,} the dotted lines represent the initial solution at $t=0$. The third column shows temperature profiles in mixture space, point symbols along the dotted line visualize the number of grid points across the initial interface. 
We observe large differences between FC and QC formulations on the coarsest grid, Figure~\ref{fig:1D}(a-c). The FC method shows unphysical velocity oscillations, whereas the QC method yields smooth profiles. Note that physical diffusion causes a change in velocity on the right side of the advected contact discontinuity. The QC method shows much higher temperatures on the dodecane side (left) compared to the FC method. 
With increasing grid resolution spurious oscillations of the FC method become less severe and eventually disappear, and the temperature profile of the QC method converges towards the FC solution. We conclude from these results that energy conservation errors - necessary to maintain velocity and pressure equilibria at interfaces without the generation of spurious oscillations - translate into errors in temperature on coarse grids and both methods converge to the same solution on sufficiently fine grids. For typical LES grid resolutions the energy conservation error of the QC method is non-negligible. 



\section{LES of ECN Spray A}

%
%
All simulations have been performed in a rectangular domain with the overall dimensions $L_x = 56$~mm ($\sim 622 D_i$) in the streamwise and  $L_y = L_z = 28$~mm ($\sim 311 D_i$) in the lateral directions. 
An adaptive Cartesian blocking strategy with a static local coarsening/refinement is used to allow for a varying grid resolution along the spray break-up trajectory to keep computational costs tractable. The grid consists of $2766$ blocks with $7$ grid refinement levels and a total number of about $15.1$ million cells.
A velocity block profile without turbulent fluctuations is prescribed at the inflow patch. The time dependent mass flow rate is taken from  \ECNmassflowrate ~with the following input parameters: injection pressure: 150~MPa; outlet diameter: $D_i = 0.09$~mm; fuel density: 703.82 kg/m$^3$; back pressure: 6~MPa; discharge coefficient: 0.89; injection time: 1.5 ms. At the outlet we prescribe the static pressure of $6$~MPa. All walls are  adiabatic.

\begin{figure*}
\centering
\includegraphics[scale=0.9]{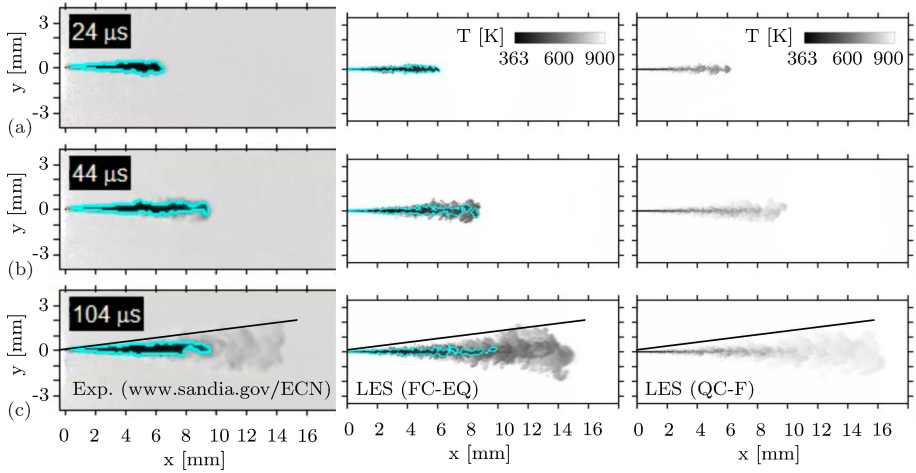}
\caption{Temporal sequence of the injection event. 
Left column: 
experimental data of \citet{Pickett:wy};  
center column: LES with FC-EQ; 
right column: LES with QC-F. 
Liquid penetration length is illustrated by a $LVF=0.15\%$ iso-contour.}
\label{fig:DiffusiveInjection}
\end{figure*}

%
%

In the following we use experimental reference data to evaluate our numerical results obtained with the quasi conservative frozen single-phase model (QC-F) and with the fully conservative equilibrium two-phase model (FC-EQ). 
The fully conservative single-phase method (FC-F) encountered numerical instabilities during the start-up phase when the jet accelerates from $0$ to $600$~m/s in just $10~\mu\mathrm{s}$. A total time interval of $1.5$~ms has been simulated. Figure~\ref{fig:DiffusiveInjection} depicts a temporal sequence of the early jet evolution ($24\mu$s-$104\mu$s). The left column shows experimental data (diffused back illumination). Center and right columns show snapshots of the temperature distribution for LES with FC-EQ and QC-F methods, respectively. In case of FC-EQ, the liquid penetration length is illustrated by the cyan iso-contour of the liquid volume fraction $LVF=0.15\%$.  
We observe a very good qualitative agreement between experimental data and LES with the FC-EQ method.  At $24~\mu$s the liquid dodecane jet extends about $6$mm into the nitrogen atmosphere, at about $44~\mu$s the liquid length has reached its quasi-steady mean. Later points in time illustrate the vapor evolution. 
QC-F and FC-EQ simulations predict a very similar vapor penetration trajectory, however, significant differences are observed for the temperature field. The dense dodecane jet heats up much quicker and mixing takes place at much higher temperatures with the QC-F model. This effect is not caused by the thermodynamic modeling approach (assumed single-phase vs.~two-phase), but rather by  energy conservation errors of the QC method.
\begin{figure*}
\centering
\includegraphics[scale=0.8]{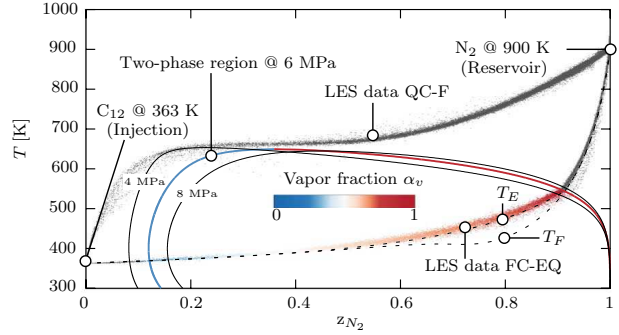}
\caption{Temperature-composition diagram for a $\text{N}_2-\text{C}_{12}\text{H}_{26}$ mixture with frozen $(T_F)$ and equilibrium $(T_E)$ mixing temperature. Scattered data depict the thermodynamic states that are obtained in the QC-F and FC-EQ LES at $144~\mu$s. For FC-EQ, points within the two-phase region are colored by vapor volume fraction.}
\label{fig:Scatter_FC-VLE-QC-F}
\end{figure*}
Figure~\ref{fig:Scatter_FC-VLE-QC-F} shows a temperature-composition phase diagram for the nitrogen-dodecane mixture together with frozen $(T_F)$ and equilibrium $(T_E)$ mixing temperature. The two-phase region is indicated at a pressure of $6$~MPa (nominal operating pressure), $4$~MPa and $8$~MPa. Scattered data depict the thermodynamic states that are obtained in the LES with the methods FC-EQ and QC-F, instantaneous data is taken from Figure~\ref{fig:DiffusiveInjection}(c). In case of FC-EQ, data points within the two-phase region are colored by the vapor volume fraction from blue to red shades. While the FC-EQ LES follows closely the equilibrium mixing temperature, we observe a completely different mixing for the QC-F LES. 
We have previously shown that the QC-F temperature prediction will eventually converge towards FC solution within the single-phase region when increasing the number of cells, i.e., reducing the energy conservation error. We therefore conclude that, for the present application and typical LES grid resolutions, the energy conservation error of the QC method 
is not controllable.

\begin{figure*}
\centering
\includegraphics[scale=0.8]{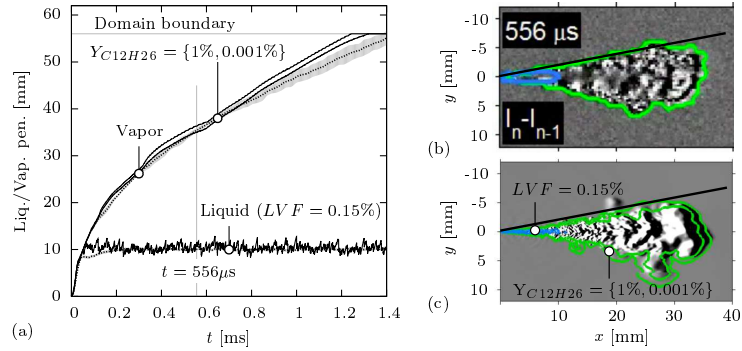}
\caption{(a)~Numerical~\gnult{1} and experimental~\gnult{4} liquid and vapor penetration trajectories. 
(b)~Experimental schlieren image. (c)~Numerical schlieren image for FC-EQ LES. 
}
\label{fig:Liq-Vap-Schlieren}
\end{figure*}

A quantitative comparison between experiment and the FC-EQ LES is given in Figure~\ref{fig:Liq-Vap-Schlieren}(a) showing 
liquid and vapor penetration trajectories. In the LES the liquid core length is defined as $L_l = \max\{x(LVF=0.15\%)\}$, vapor penetration $L_v$ is shown for the definitions $\max\{x(Y_{C12H26}=1\%)\}$ and $\max\{x(Y_{C12H26}=0.001\%)\}$.  We observe an excellent agreement of $L_l$ with the experimental time-resolved signal. It is important to note that the measured $L_l$ depends on the chosen threshold value. Based on a thorough analysis based on Mie-scatter theory together with assumptions on droplet diameters, \citet{Pickett:wy} conclude that the $LVF$ threshold representing their liquid length is expected to be less than 0.15\% at Spray A conditions. The experimental time-averaged liquid length fluctuates by approximately $\pm 1$mm about the quasi-steady mean of $10.4$mm; this value is in excellent agreement with our LES data for the threshold value of $0.15\%$. In order to evaluate the sensitivity on the threshold value we computed $L_l$ for $LVF = \{ 3\%, 1\%, 0.15\%, 0.05\% \}$  and obtained $L_l = \{ 8.83, 9.91, 10.40, 10.49 \}$mm, respectively.

We also observe a good agreement of the vapor penetration trajectory up to $\sim 0.8$ms. At later times the penetration depth is slightly over predicted. 
In the experiment, the vapor penetration length is derived from high speed schlieren images. Figure~\ref{fig:Liq-Vap-Schlieren}(b-c) give an impression on how a mixture fraction threshold compares to a schlieren image.  The numerical schlieren image shows the axial density gradient $\partial \rho / \partial x$ spatially averaged along the z-direction. 
Numerical and experimental image are strikingly similar. Quantitatively, a definition of the vapor penetration depth by a $1\%$ mixture fraction threshold seems to slightly under predict the vapor penetration derived from a schlieren image, mainly in the long term evolution. We therefore do not recommend to track values larger $1\%$.

%


\begin{figure*}
\centering
\includegraphics[scale=0.8]{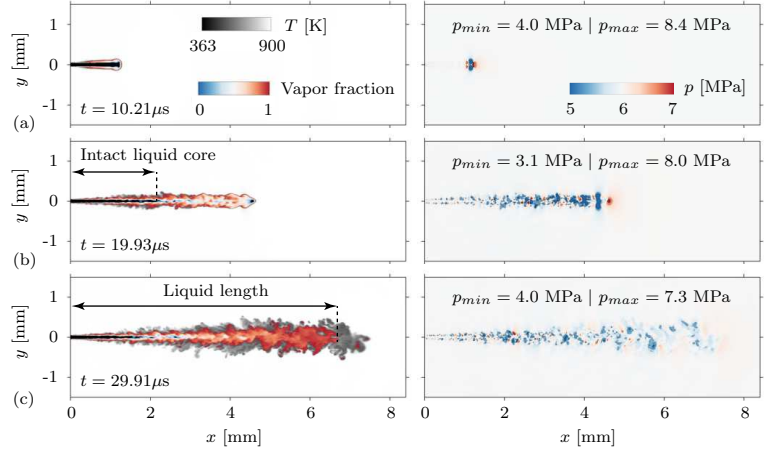}
\caption{Temporal sequence of temperature (left) and pressure (right) for FC-EQ LES.}
\label{fig:Near-nozzle}
\end{figure*}

Figure~\ref{fig:Near-nozzle} shows a temporal sequence of the spray structure in the near-nozzle field at a very early state, $10~\mu\mathrm{s}$, $20~\mu\mathrm{s}$ and $30~\mu\mathrm{s}$ after injection start. In the left column we show instantaneous snapshots of the temperature field (contour levels are shown for \mbox{$363~\mathrm{K} < T < 900~\mathrm{K}$}, dark to light grey shades). Superimposed is the vapor volume fraction distribution (blue to red shades) for the two-phase region within which the isochoric-isoenergetic flash problem was solved. Contours of the corresponding pressure fields \mbox{($5~\mathrm{MPa} < p < 7~\mathrm{MPa}$}, from blue to red shades) are shown in the right column. We see that the dodecane-nitrogen mixture locally experiences pressures much different from the average ambient pressure. A very low pressure ($\sim 3$~MPa) can be observed at the tip of the jet due to the start-up vortex ring. Even in the fully developed steady state we see pressure fluctuations in the shear layer in the order of $\pm 10~\mathrm{bar}$. 

We mentioned above that we were not able to simulate Spray A with a conservative single-phase model (FC-F). 
The encountered instabilities are caused by our single-phase thermodynamics, which yields ill-defined states at low pressures that occur in well resolved vortex cores. Our fully conservative two-phase LES model (FC-EQ) did not face any stability problems because the more sophisticated model can resolve coexisting subcritical two-phase states, thus avoiding unphysical states.

\section{Conclusion}

A detailed multi-species two-phase thermodynamic equilibrium model for the Eulerian LES of turbulent mixing at high pressures has been presented and applied for LES of liquid-fuel injection at transcritical operating conditions. The thermodynamics model is based on cubic equations of state and vapor-liquid equilibrium calculations. It can thus accurately represent supercritical states as well as coexisting multi-component subcritical two-phase states. Computational results for the transcritical dodecane injection ECN Spray A  case demonstrate the excellent performance of the model. We saw that the Spray A dodecane-nitrogen mixture locally experiences pressures significantly below the nominal operating pressure of $6$~MPa when the jet accelerates from $0$ to $600$~m/s in just $10~\mu\mathrm{s}$. For these harsh conditions LES with a conservative dense-gas single-phase approach exhibit large spurious pressure oscillations that may cause numerical instability even with low-order upwind numerics. It has been previously suggested that stable time integration of single-phase thermodynamic models can be obtained by ''energy-correction methods'' that sacrifice energy conservation in some way. We therefore compared a fully conservative formulation of the governing equations with a quasi conservative formulation based on a pressure evolution equation. A one-dimensional multi-component advection-diffusion test cases proved physical and numerical consistency of both methods and convergence towards the same solution for sufficiently fine grids. On coarser grids, however, energy conservation errors associated with the quasi conservative formulation caused a significant over-prediction of the temperature. LES with our new fully conservative multi-component two-phase equilibrium model did not show any stability problems and yield numerical predictions that are in very good agreement with available experimental data.

%
%

\subsection*{Acknowledgments}
We are grateful to Chao Ma, Daniel Banuti, Lluis Jofre, Matthias Ihme and Laurent Selle for valuable discussions during the summer program.
This project was partially funded by the German Research Foundation (DFG) through the SFB TRR-40 and the TUM Graduate School. We would also like to thank the Gauss Centre for Supercomputing e.V. for providing computing time on SuperMUC at Leibniz Supercomputing Centre (LRZ).


\end{document}